# Momentum sharing in imbalanced Fermi systems

**Authors:** O. Hen[40]*, M. Sargsian[10], L.B. Weinstein[27], E. Piasetzky[40], H. Hakobyan[34,39], D. W. Higinbotham[33], M. Braverman[40], W.K. Brooks[34], S. Gilad[41], K. P. Adhikari[27], J. Arrington[1], G. Asryan[39], H. Avakian[33], J. Ball[7], N. A. Baltzell[1], M. Battaglieri[17], A. Beck[40,43], S. May-Tal Beck[40,43], I. Bedlinskiy[20], W. Bertozzi[41], A. Biselli[42], V. D. Burkert[33], T. Cao[32], D. S. Carman[33], A. Celentano[17], S. Chandavar[26], L. Colaneri[18], P. L. Cole[14,6,33], V. Crede[11], A. D'Angelo[18,30], R. De Vita[17], A. Deur[33], C. Djalali[32], D. Doughty[8,33], M. Dugger[2], R. Dupre[19], H. Egiyan[33], A. El Alaoui[1], L. El Fassi[27], L. Elouadrhiri[33], G. Fedotov[32,31], S. Fegan[17], T. Forest[14], B. Garillon[19], M. Garcon[7], N. Gevorgyan[39], Y. Ghandilyan[39], G. P. Gilfoyle[29], F. X. Girod[33], J. T. Goetz[26], R. W. Gothe[32], K. A. Griffioen[38], M. Guidal[19], L. Guo[10,33], K. Hafidi[1], C. Hanretty[37], M. Hattawy[19], K. Hicks[26], M. Holtrop[24], C. E. Hyde[27], Y. Ilieva[32,13], D. G. Ireland[36], B.I. Ishkanov[31], E. L. Isupov[31], H. Jiang[32], H. S. Jo[19], K. Joo[9], D. Keller[37], M. Khandaker[14,25], A. Kim[22], W. Kim[22], F. J. Klein[6], S. Koirala[27], I. Korover[40], S. E. Kuhn[27], V. Kubarovsky[33], P. Lenisa[15], W. I. Levine[5], K. Livingston[36], M. Lowry[33], H. Y. Lu[32], I. J. D. MacGregor[36], N. Markov[9], M. Mayer[27], B. McKinnon[36], T. Mineeva[9], V. Mokeev[19,33], A. Movsisyan[15], C. Munoz Camacho[19], B. Mustapha[1], P. Nadel-Turonski[33], S. Niccolai[19], G. Niculescu[21], I. Niculescu[21], M. Osipenko[17], L. L. Pappalardo[15], R. Paremuzyan[39], K. Park[33,22], E. Pasyuk[33], W. Phelps[10], S. Pisano[16], O. Pogorelko[20], J. W. Price[3], S. Procureur[7], Y. Prok[27,37], D. Protopopescu[36], A. J. R. Puckett[9], D. Rimal[10], M. Ripani[17], B. G. Ritchie[2], A. Rizzo[18], G. Rosner[36], P. Rossi[33], P. Roy[11], F. Sabati´e[7], D. Schott[13], R. A. Schumacher[5], Y. G. Sharabian[33], G. D. Smith[35], R. Shneor[40], D. Sokhan[36], S. S. Stepanyan[22], S. Stepanyan[33], P. Stoler[28], S. Strauch[32,13], V. Sytnik[34], M. Taiuti[12], S. Tkachenko[37], M. Ungaro[33], A. V. Vlassov[20], E. Voutier[23], N. K. Walford[6], X. Wei[33], M. H. Wood[4,32], S. A. Wood[33], N. Zachariou[32], L. Zana[35,24], Z. W. Zhao[37], X. Zheng[37], and I. Zonta[18].

(Jefferson Lab CLAS Collaboration)

**Affiliations:**

[1]Argonne National Laboratory, Argonne, Illinois 60439.

[2]Arizona State University, Tempe, Arizona 85287-1504.

[3]California State University, Dominguez Hills, Carson, CA 90747.

[4]Canisius College, Buffalo, NY.

[5]Carnegie Mellon University, Pittsburgh, Pennsylvania 15213.

[6]Catholic University of America, Washington, D.C. 20064.

[7]CEA, Centre de Saclay, Irfu/Service de Physique Nucl´eaire, 91191 Gif-sur-Yvette, France.

[8]Christopher Newport University, Newport News, Virginia 23606.

[9]University of Connecticut, Storrs, Connecticut 06269.

[10]Florida International University, Miami, Florida 33199.

[11]Florida State University, Tallahassee, Florida 32306.

[12]Universita` di Genova, 16146 Genova, Italy.

[13]The George Washington University, Washington, DC 20052.

[14]Idaho State University, Pocatello, Idaho 83209.

[15]INFN, Sezione di Ferrara, 44100 Ferrara, Italy.

[16]INFN, Laboratori Nazionali di Frascati, 00044 Frascati, Italy.

[17]INFN, Sezione di Genova, 16146 Genova, Italy.

[18]INFN, Sezione di Roma Tor Vergata, 00133 Rome, Italy.

[19]Institut de Physique Nucl´eaire ORSAY, Orsay, France.

[20]Institute of Theoretical and Experimental Physics, Moscow, 117259, Russia.

[21]James Madison University, Harrisonburg, Virginia 22807.

[22]Kyungpook National University, Daegu 702-701, Republic of Korea.

[23]LPSC, Universite Joseph Fourier, CNRS/IN2P3, INPG, Grenoble, France.

[24]University of New Hampshire, Durham, New Hampshire 03824-3568.

[25]Norfolk State University, Norfolk, Virginia 23504.

[26]Ohio University, Athens, Ohio 45701.

[27]Old Dominion University, Norfolk, Virginia 23529.

[28]Rensselaer Polytechnic Institute, Troy, New York 12180-3590.

[29]University of Richmond, Richmond, Virginia 23173.

[30]Universita' di Roma Tor Vergata, 00133 Rome Italy.

[31]Skobeltsyn Institute of Nuclear Physics, Lomonosov Moscow State University, 119234 Moscow, Russia.

[32]University of South Carolina, Columbia, South Carolina 29208.

[33]Thomas Jefferson National Accelerator Facility, Newport News, Virginia 23606.

[34]Universidad T´ecnica Federico Santa Mar´ıa, Casilla 110-V Valpara´ıso, Chile.

[35]Edinburgh University, Edinburgh EH9 3JZ, United Kingdom.

[36]University of Glasgow, Glasgow G12 8QQ, United Kingdom.

[37]University of Virginia, Charlottesville, Virginia 22901.

[38]College of William and Mary, Williamsburg, Virginia 23187-8795.

[39]Yerevan Physics Institute, 375036 Yerevan, Armenia.

[40]Tel Aviv University, Tel Aviv 69978, Israel.

[41]Massachusetts Institute of Technology, Cambridge, MA 02139.

[42]Fairfield University, Fairfield, CT 06824.

[43]NRCN, P.O. Box 9001, Beer-Sheva 84190, Israel.

*Correspondence to: Or Hen (or.chen@mail.huji.ac.il).

**Abstract**: The atomic nucleus is composed of two different kinds of fermions, protons and neutrons. If the protons and neutrons did not interact, the Pauli exclusion principle would force the majority fermions (usually neutrons) to have a higher average momentum. Our high-energy electron scattering measurements using $^{12}C$, $^{27}Al$, $^{56}Fe$ and $^{208}Pb$ targets show that, even in heavy neutron-rich nuclei, short-range interactions between the fermions form correlated high-momentum neutron-proton pairs. Thus, in neutron-rich nuclei, protons have a greater probability than neutrons to have momentum greater than the Fermi momentum. This finding has implications ranging from nuclear few body systems to neutron stars and may also be observable experimentally in two-spin state, ultra-cold atomic gas systems.

**Main Text:** Many-body systems composed of interacting fermions are common in nature, ranging from high-temperature superconductors and Fermi liquids to atomic nuclei, quark matter and neutron-stars. Particularly intriguing are systems that include a short-range interaction that is strong between unlike fermions and weak between fermions of the same kind. Recent theoretical advances show that even though the underlying interaction can be very different, these systems share several universal features (*1-4*). In all these systems, this interaction creates short-range correlated (SRC) pairs of unlike fermions with a large relative momentum ($k_{rel} > k_F$) and a small center of mass (CM) momentum ($k_{tot} < k_F$), where $k_F$ is the Fermi momentum of the system. This pushes fermions from low momenta ($k < k_F$ where k is the fermion momentum) to high momenta ($k > k_F$), creating a "high momentum tail".

SRC pairs in atomic nuclei have been studied using many different reactions, including pickup, stripping and electron and proton scattering. The results of these studies highlighted the importance of correlations in nuclei, which lead to a high momentum tail and decreased occupancy of low-lying nuclear states (*5-13*).

Recent experimental studies of balanced (symmetric) interacting Fermi systems, with an equal number of fermions of the two kinds, confirmed these predictions of a high momentum tail populated almost exclusively by pairs of unlike fermions (*8-11,14-16*). These experiments were done using very different Fermi systems: protons and neutrons in atomic nuclei and two-spin state ultra-cold atomic gasses. These systems span more than 15 orders of magnitude in Fermi energy from $10^6$ to $10^{-9}$ eV and exhibit different short-range interactions (predominantly a strong tensor interaction in the nuclear systems (*8,9,17,18*), and a tunable Feshbach resonance in the atomic system (*14,15*)). For cold atoms Ref. (*1-3*) showed that the momentum density decreases as $C/k^4$ for large k. The scale factor, C, is known as Tan's contact and describes many properties of the system (*4*). Similar pairing of nucleons in nuclei with $k > k_F$ was also predicted in (*19*).

Here, we extend these previous studies to imbalanced (asymmetric) nuclear systems, with unequal numbers of the different fermions. When there is no interaction, the Pauli exclusion principle pushes the majority fermions (usually neutrons) to a higher average momentum. Including a short-range interaction introduces a new universal feature: the probability for a fermion to have momentum $k > k_F$ is greater for the minority than for the majority fermions. This is because the short-range interaction populates the high-momentum tail with equal numbers of majority and minority fermions, thereby leaving a larger fraction of majority fermions in low momentum states ($k<k_F$), see Fig. 1. In neutron-rich nuclei this increases the average proton momentum and may even result in protons having higher average momentum than neutrons, inverting the momentum sharing in imbalanced nuclei from that in non-interacting systems. Theoretically, this can happen because of the tensor part of the nucleon-nucleon interaction, which creates predominantly spin-1 isospin-0 neutron-proton (np) SRC pairs (*17,18*).

Here, we identify SRC pairs in the high-momentum tail of nuclei heavier than carbon with more neutrons than protons (i.e. N>Z). The data show the universal nature of SRC pairs, which even in lead (N/Z = 126/82) are still predominantly np pairs. This np-pair dominance causes a greater fraction of protons than neutrons to have high momentum in neutron-rich nuclei.

The data presented here were collected in 2004 in Hall B of the Thomas Jefferson National Accelerator Facility using a 5.014 GeV electron beam incident on $^{12}C$, $^{27}Al$, $^{56}Fe$ and $^{208}Pb$ targets. We measured electron induced two-proton knockout reactions (Fig. 2). The CEBAF Large Acceptance Spectrometer (CLAS) (*20*) was used to detect the scattered electron and emitted protons. CLAS uses a toroidal magnetic field and six independent sets of drift chambers, time-of-flight scintillation counters, Cerenkov counters, and electro-magnetic calorimeters for charged particle identification and trajectory reconstruction (Fig. 2) (*16*).

We selected events in which the electron interacts with a single fast proton from an SRC pair in the nucleus (*9,16*) by requiring a large four-momentum transfer $Q^2 = \vec{q}^2 - (\omega / c)^2 > 1.5$ GeV$^2$/c$^2$, where $\vec{q}$ and $\omega$ are the three-momentum and energy transferred to the nucleus respectively; and Bjorken scaling parameter $x_B = Q^2/(2m_N \cdot \omega / c) > 1.2$, where $m_N$ is the nucleon mass. To ensure selection of events in which the knocked-out proton belonged to an SRC pair, we further required missing momentum $300 < |\vec{p}_{miss}| < 600$ MeV/c, where $\vec{p}_{miss} = \vec{p}_p - \vec{q}$ with $\vec{p}_p$ the measured proton momentum. We suppressed contributions from inelastic excitations of the struck nucleon by limiting the reconstructed missing mass of the two-nucleon system $m_{miss} < 1.1$ GeV/c$^2$. In each event, the leading proton that absorbed the transferred momentum was identified by requiring that its momentum $\vec{p}_p$ is within 25$^{\circ}$ of $\vec{q}$ and that $|\vec{p}_p|/|\vec{q}| \geq 0.6$ (*16,21*).

When a second proton was detected with momentum greater than 350 MeV/c, it was emitted almost diametrically opposite to $\vec{p}_{miss}$, see Fig. S-19. The observed backward-peaked angular distributions are very similar for all four measured nuclei. This backward peak is a strong signature of SRC pairs, indicating that the two emitted protons were largely back-to-back in the initial state, having a large relative momentum and a small center-of-mass momentum (*8,9*). This is a direct observation of proton-proton (pp) SRC pairs in a nucleus heavier than $^{12}$C.

Electron scattering from high missing momentum protons is dominated by scattering from protons in SRC pairs (*9*). The measured single-proton knockout (e,e'p) cross section is sensitive to the number of pp and np SRC pairs in the nucleus, while the two-proton knockout (e,e'pp) cross section is only sensitive to the number of pp-SRC pairs. Very few of the single-proton knockout events also contained a second proton, therefore there are very few pp pairs and the knocked out protons predominantly originated from np pairs.

To quantify this, we extracted the [A(e,e'pp)/A(e,e'p)] / [$^{12}$C(e,e'pp)/$^{12}$C(e,e'p)] cross section double ratio for nucleus *A* relative to $^{12}$C. The double ratio is sensitive to the ratio of np to pp SRC pairs in the two nuclei (*16*). Previous measurements have shown that in $^{12}$C nearly every high-momentum proton (k > 300 MeV/c > $k_F$) has a correlated partner nucleon, with np pairs outnumbering pp pairs by a factor of ~20 (*8,9*).

To estimate the effects of final-state interactions (re-interaction of the outgoing nucleons in the nucleus), we calculated attenuation factors for the outgoing protons and the probability of the electron scattering from a neutron in an np pair, followed by a neutron-proton single charge exchange (SCX) reaction leading to two outgoing protons. These correction factors are calculated as in Ref. (*9*) using the Glauber approximation (*22*) with effective cross sections that reproduce previously measured proton transparencies (*23*) and using the measured SCX cross-section of Ref. (*24*).

We extracted the cross section ratios and deduced the relative pair fractions from the measured yields following Ref. (*21*), see (*16*) for details.

Figure 3 shows the extracted fractions of np and pp SRC pairs from the sum of pp and np pairs in nuclei, including all statistical, systematic and model uncertainties. Our measurements are not sensitive to neutron-neutron SRC pairs. However, by a simple combinatoric argument, even in $^{208}$Pb these would only be $(N/Z)^2 \sim 2$ times the number of pp pairs. Thus, np-SRC pairs dominate in all measured nuclei, including neutron-rich imbalanced ones.

The observed dominance of np-over-pp pairs implies that even in heavy nuclei, SRC pairs are dominantly in a spin triplet state (spin 1 isospin 0), which are a consequence of the tensor part of the nucleon-nucleon interaction (*17,18*). It also implies there are as many high-momentum protons as neutrons (Fig. 1) so that the fraction of protons above the Fermi momentum is greater than that of neutrons in neutron-rich nuclei (*25*).

In light ($A \leq 12$) imbalanced nuclei Variational Monte Carlo (VMC) calculations (*26*) show that this results in a greater average momentum for the minority component, see Table S-1. The minority component can also have a greater average momentum in heavy nuclei if the Fermi momenta of protons and neutrons are not too dissimilar. For heavy nuclei, an np-dominance toy model that quantitatively describes the features of the momentum distribution shown in Fig. 1 shows that in imbalanced nuclei the average proton kinetic energy is greater than that of the neutron, up to ~20% in $^{208}$Pb (*16*).

The observed np-dominance of SRC pairs in heavy imbalanced nuclei may have wide-ranging implications. Neutrino scattering from two nucleon currents and SRC pairs is important for the analysis of neutrino-nucleus reactions, which are used to study the nature of the electro-weak interaction (*27-29*). In particle physics, the distribution of quarks in these high-momentum nucleons in SRC-pairs might be modified from that of free nucleons (*30,31*). Because each proton has a greater probability to be in an SRC-pair than a neutron and the proton has two u quarks for each d quark, the u-quark distribution modification could be greater than that of the d quarks (*19,30*). This could explain the difference between the weak mixing angle measured on an iron target by the NuTeV experiment and the Standard Model value (*32-34*).

In astrophysics, the nuclear symmetry-energy is important for various systems, including neutron stars, the neutronization of matter in core-collapse supernovae, and *r*-process nucleosynthesis (*35*). The decomposition of the symmetry energy at saturation density ($\rho_0 \approx 0.17$ fm$^{-3}$, the maximum density of normal nuclei) into its kinetic and potential parts and its value at supra-nuclear densities ($\rho > \rho_0$), are not well constrained, largely because of the uncertainties in the tensor component of the nucleon-nucleon interaction (*36-39*). Although at supra-nuclear densities other effects are relevant, the inclusion of high-momentum tails, dominated by tensor-force induced np-SRC pairs, can notably soften the nuclear symmetry energy (*36-39*). Our measurements of np-SRC pair dominance in heavy imbalanced nuclei can help constrain the nuclear aspects of these calculations at saturation density.

Based on our results in the nuclear system, we suggest extending the previous measurements of Tan's contact in balanced ultra-cold atomic gases to imbalanced systems where the number of atoms in the two spin states is different. The large experimental flexibility of these systems will allow observing dependence of the momentum sharing inversion on the asymmetry, density, and the strength of the short-range interaction.

**Acknowledgments:** This work was supported by: US Department of Energy and National Science Foundation, Israel Science Foundation, Chilean Comisión Nacional de Investigación Científica y Tecnológica, French Centre National de la Recherche Scientifique and Commissariat a l'Energie Atomique, French-American Cultural Exchange, Italian Istituto Nazionale di Fisica Nucleare, National Research Foundation of Korea, and United Kingdom's Science and Technology Facilities Council.

Jefferson Science Associates operates the Thomas Jefferson National Accelerator Facility for the United States Department of Energy, Office of Science, Office of Nuclear Physics under contract DE-AC05-06OR23177. The raw data from this experiment are archived in Jefferson Lab's mass storage silo.

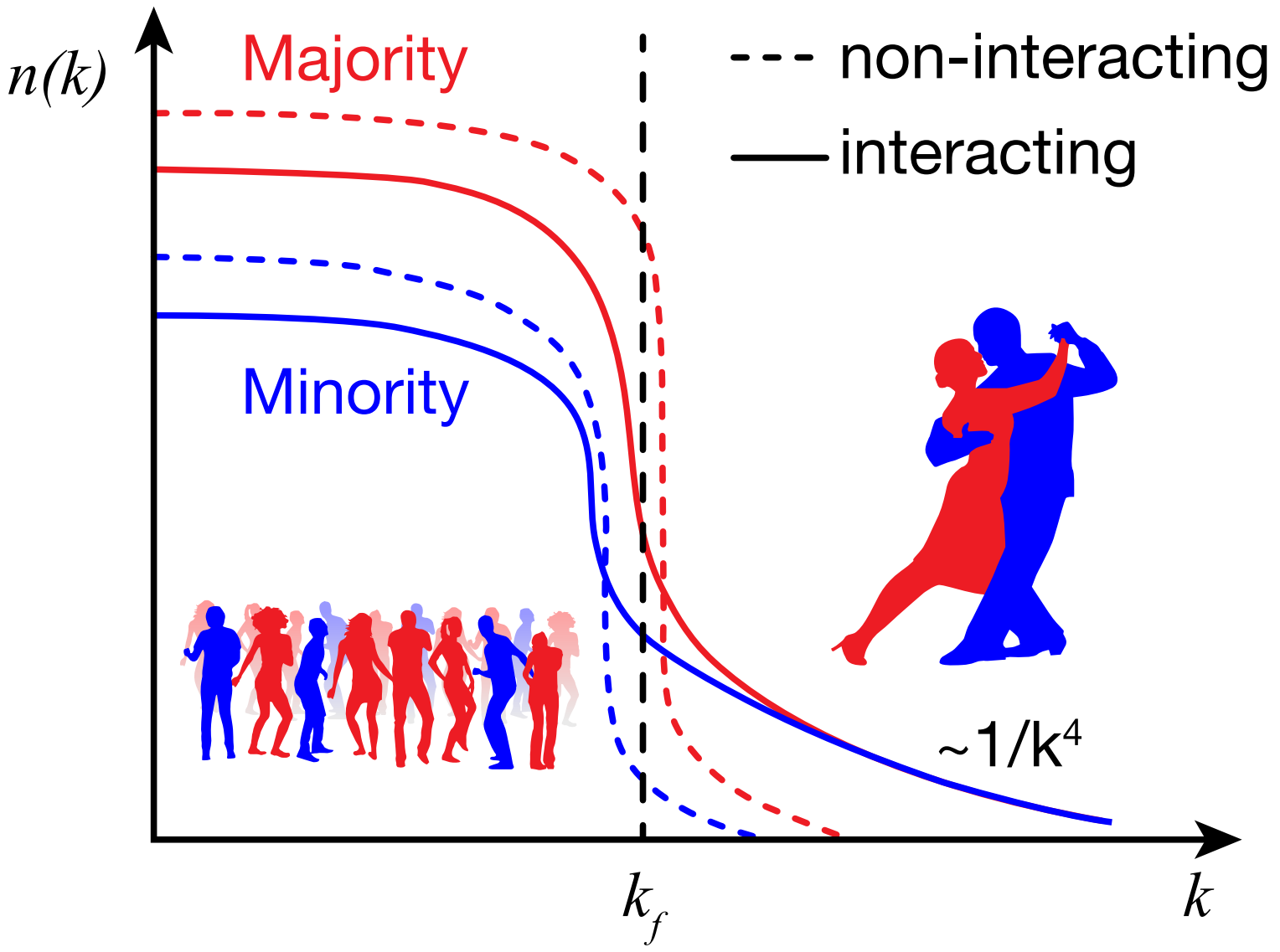

**Fig. 1**. A schematic representation of the momentum distribution, n(k), of two-component imbalanced Fermi systems. The dashed lines show the non-interacting system whereas the solid lines show the effect of including a short-range interaction between different fermions. Such interactions create a high-momentum ($k>k_F$ where $k_F$ is the Fermi momentum of the system) tail. This is analogous to a dance party with a majority of girls, where boy-girl interactions will make the average boy dance more than the average girl.

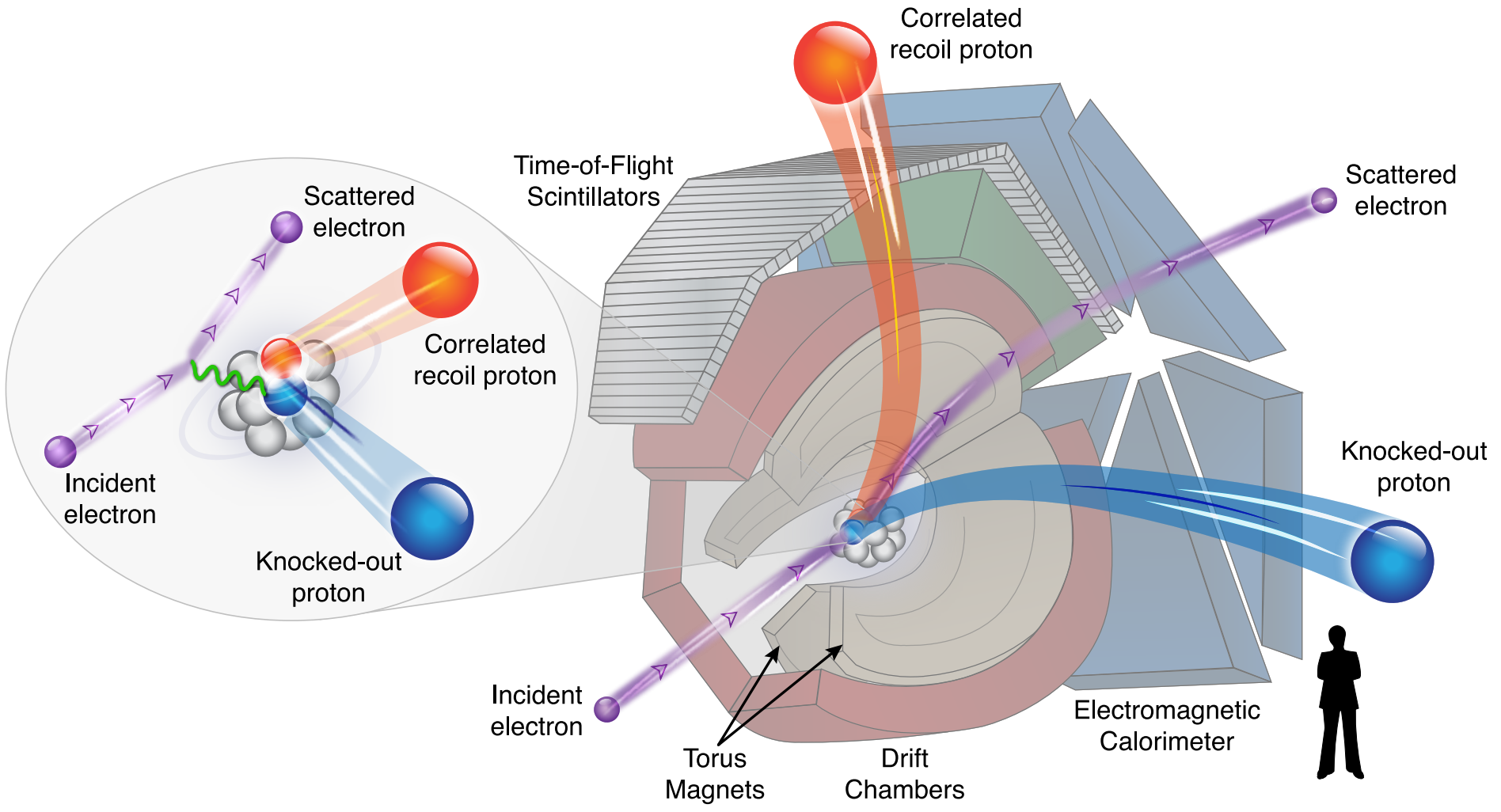

**Fig. 2.** Illustration of the CLAS detector with a reconstructed two-proton knockout event. For clarity, not all CLAS detectors and sectors are shown. The inset shows the reaction in which an incident electron scatters from a proton-proton pair via the exchange of a virtual photon. The human figure is shown for scale.

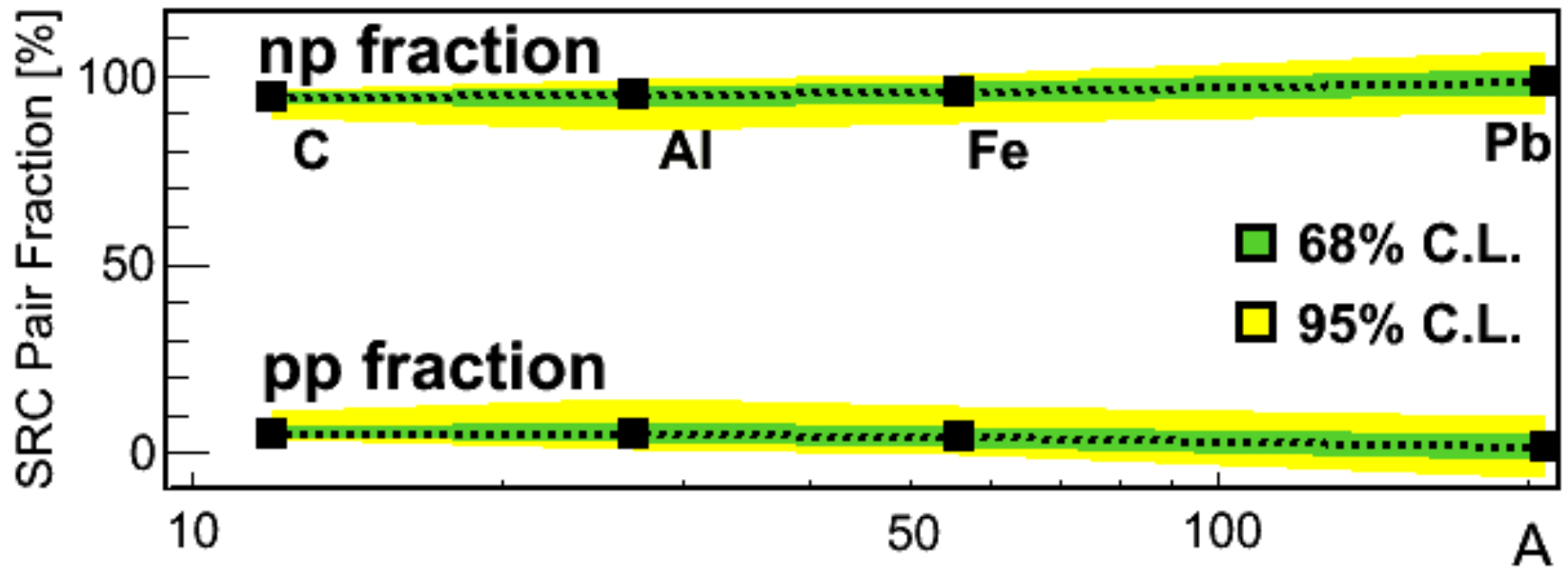

**Fig. 3.** The extracted fractions of np (top) and pp (bottom) SRC pairs from the sum of pp and np pairs in nuclei. The green and yellow bands reflect 68% and 95% confidence levels, respectively (9). np-SRC pairs dominate over pp-SRC pairs in all measured nuclei.

**Supplementary Materials:**
Materials and Methods
Figures S1 – S32
Tables S1-S8
References (*40-51*)